\documentclass[twocolumn,superscriptaddress,showpacs,preprintnumbers,amsmath,amssymb,aps,pre]{revtex4}
\usepackage{graphicx}
\usepackage{dcolumn}
\usepackage{bm}
\usepackage{color}

\begin{document}
\title{Fluctuation theorem and natural time analysis}
\author{N. V. Sarlis}
\affiliation{Solid State Section and Solid Earth Physics
Institute, Physics Department, University of Athens,
Panepistimiopolis, Zografos 157 84, Athens, Greece}
\author{P. A. Varotsos}
\email{pvaro@otenet.gr} \affiliation{Solid State Section and Solid
Earth Physics Institute, Physics Department, University of Athens,
Panepistimiopolis, Zografos 157 84, Athens, Greece}
\author{E. S. Skordas}
\affiliation{Solid State Section and Solid Earth Physics
Institute, Physics Department, University of Athens,
Panepistimiopolis, Zografos 157 84, Athens, Greece}

\begin{abstract}
Upon employing a natural time window of fixed length sliding
through a time series, an explicit interrelation between the
variability $\beta$ of the variance $\kappa_1$($=\langle
\chi^2\rangle - \langle \chi \rangle^2$) of natural time $\chi$
and events' correlations is obtained. In addition, we investigate
the application of the fluctuation theorem, which is a general
result for systems far from equilibrium, to the variability
$\beta$. We consider for example,  major earthquakes that are
nonequilibrium critical phenomena. We find that four (out of five)
mainshocks in California during 1979-2003 were preceded by $\beta$
minima lower than the relative thresholds deduced from the
fluctuation theorem, thus signalling an impending major event.
\end{abstract} 
 \maketitle

Entropy production is a measure of the irreversibility of a
thermodynamic process: the difficulty, even impossibility, of
reversing the observed often macroscopic behavior of a system that
exchanges heat or matter with a complex environment (e.g., see
Ref.\cite{FORD12} and references therein). The breakage of time
reversal symmetry associated with thermodynamic irreversibility
has focused enormous discussion for more than a century. Despite
of such concerns, however, the concept of entropy generation in
the thermodynamics of large systems has been applied widely. From
microscopic point of view, efforts towards understanding the
nature of the entropy and its production -mainly focused on the
one way character of the second law- have been attempted. They
modelled the microscopic evolution of a system and its environment
in the frame of stochastic dynamics \cite{GARD09} and stochastic
thermodynamics \cite{SEIF05,SEIF08,SEKI10}, but interpretations
based on deterministic dynamics (e.g., see Ref.\cite{EVA11A}) were
also forwarded.

An intense interest towards the latter interpretations has been
renewed when Evans, Cohen and Morris in 1993 considered the
fluctuations of the entropy production rate in a shearing fluid,
and proposed the so-called fluctuation relation or the first
fluctuation theorem \cite{EVA93}. This is considered \cite{MARC08}
to represent a general result concerning systems arbitrarily far
from equilibrium. The proof of the fluctuation \cite{EVA94} and
related theorems \cite{EVA02} shows how irreversible macroscopic
behavior arises from time reversible microscopic equations of
motion. The two theoretical results that illustrate this clearly
are the second law inequality \cite{EVA04} and the very recent
mechanical proof \cite{EVA11B} of Clausius' inequality without the
prior assumption of the second ``law'' of thermodynamics. These
two results have been obtained without treating the nonequilibrium
entropy, but used instead a quantity termed dissipation function
first defined \cite{SEAR00} in 2000. On the basis of this
function, being a path function and not a state function, the
relaxation of a system to equilibrium, which is inherently a
nonequilibrium process, can be quantified \cite{REID12}.

It has been emphasized in Ref. \cite{EVA11A} that, unlike linear
irreversible thermodynamics, the fluctuation and related theorems
are exact for systems of arbitrary size as well as for systems
arbitrarily near to, or far from equilibrium, as mentioned. This
is why we shall employ here the fluctuation theorem for the
purpose of the present study.

This theorem
\cite{EVA93,EVA94,EVA95,EVA96,SEAR00A,GALL95A,GALL95B} gives a
general formula for  the probability ratio that in a thermostated
dissipative system, the time average entropy production
$\bar{\Sigma}_t$ takes a value $A$ to minus the value $-A$,
\begin{equation}\label{eq1}
    Pr(\bar{\Sigma}_t / k_B=A)/Pr(\bar{\Sigma}_t / k_B=-A)=\exp[At]
\end{equation}
from which it is obvious that as the averaging time or {system
size} increases, it becomes exponentially likely that the entropy
production will be positive. The theorem was initially proposed
\cite{EVA93} for nonequilibrium steady states that are
thermostated in such a way that the total energy of the system is
constant. Subsequently, it was shown \cite{GALL95A,GALL95B} that
this theorem can be proved for sufficiently chaotic, iso-energetic
nonequilibrium systems using the Sinai-Ruelle-Bowen measure, as
well as for purely Hamiltonian systems with or without applied
dissipative fields \cite{EVA01} and for a wide class of stochastic
nonequilirium systems \cite{LOBO99,SEAR99}.

It is one of the two basic aims of this paper to investigate for
the first time the application of the fluctuation theorem to the
case of earthquakes which may be considered (e.g.
\cite{TUR97,HOL06}) as nonequilibrium critical phenomena (the
mainshock being the new phase). They exhibit complex correlations
in time, space and magnitude $M$ which have been recently studied
by several workers (e.g., see Refs.
\cite{TEN12,LEN11,LIP12,TEL10B,HUA08}). In particular, the present
investigation will be made by applying the fluctuation theorem to
the order parameter fluctuations that result from the analysis of
the time series in a new time domain termed\cite{NAT02} natural
time $\chi$. This is so, because natural time analysis allows us
to identify\cite{PNAS} when a complex system approaches a critical
point (for a review see Ref. \cite{SPRINGER}) and in addition
enables the introduction of an order parameter for seismicity. The
present study has been motivated by the following two findings
related to the variability $\beta$ (defined below) of the order
parameter of seismicity \cite{NAT05C}: First, it captures the
events' correlations, as shown here (see Appendix), which constitutes the other basic aim of
this paper. Second, the quantity $\beta$ exhibits characteristic
minima \cite{EPL12} before the occurrence of major events.

In a time series comprising $N$ earthquakes, the natural time
$\chi_k = k/N$ serves as an index for the occurrence of the $k$-th
earthquake. In natural time analysis the pair $(\chi_k, Q_k)$ is
studied, where $Q_k$ is the energy released during the $k$-th
earthquake of magnitude $M_k$. One may alternatively study the
pair $(\chi_k,p_k)$, where $ p_k={Q_k}/{\sum_{n=1}^NQ_n}$ is the
normalized energy released during the $k$-th earthquake, and $Q_k$
-and hence $p_k$- is estimated through the  relation \cite{KAN78}
$Q_k \propto 10^{1.5M_k}$. The variance $\kappa_1$($=\langle
\chi^2\rangle - \langle \chi \rangle^2$) of $\chi$ weighted for
$p_k$, is given by \cite{NAT02,NAT03B,NAT03A,NAT05C}
\begin{equation}\label{kappa1}
\kappa_1=\sum_{k=1}^N p_k (\chi_k)^2- \left(\sum_{k=1}^N p_k
\chi_k \right)^2
\end{equation}
This quantity, as shown in Ref. \cite{NAT05C}, can be also
considered as an order parameter for seismicity.

\begin{figure*}
\includegraphics{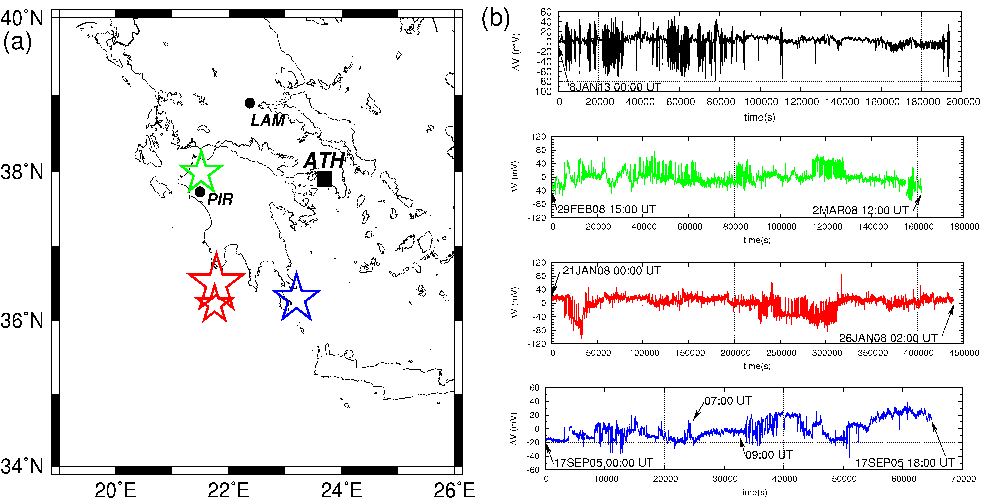}
\caption{(color online) (a) Major earthquakes in Greece on 8 June
2008 (green, magnitude $M_w=$6.4), 14 February 2008 (red,
$M_w=$6.9 and 6.4) and 8 January 2006 (blue, $M_w=$6.7). (b) Their
preceding SES activities recorded at Pirgos (PIR) measuring
station located in western Greece are shown (with the
corresponding color) in the lower three channels. Furthermore, an
SES activity initiated recently on 8 January 2013 at a station in
central Greece labelled LAM in (a) is depicted in the upper
channel of (b). Additional SES activities were recorded at LAM
from 31 March to 11 April 2013, see also Ref.\cite{neo}
}\label{f1}
\end{figure*}

\begin{figure*}
\includegraphics[scale=0.8]{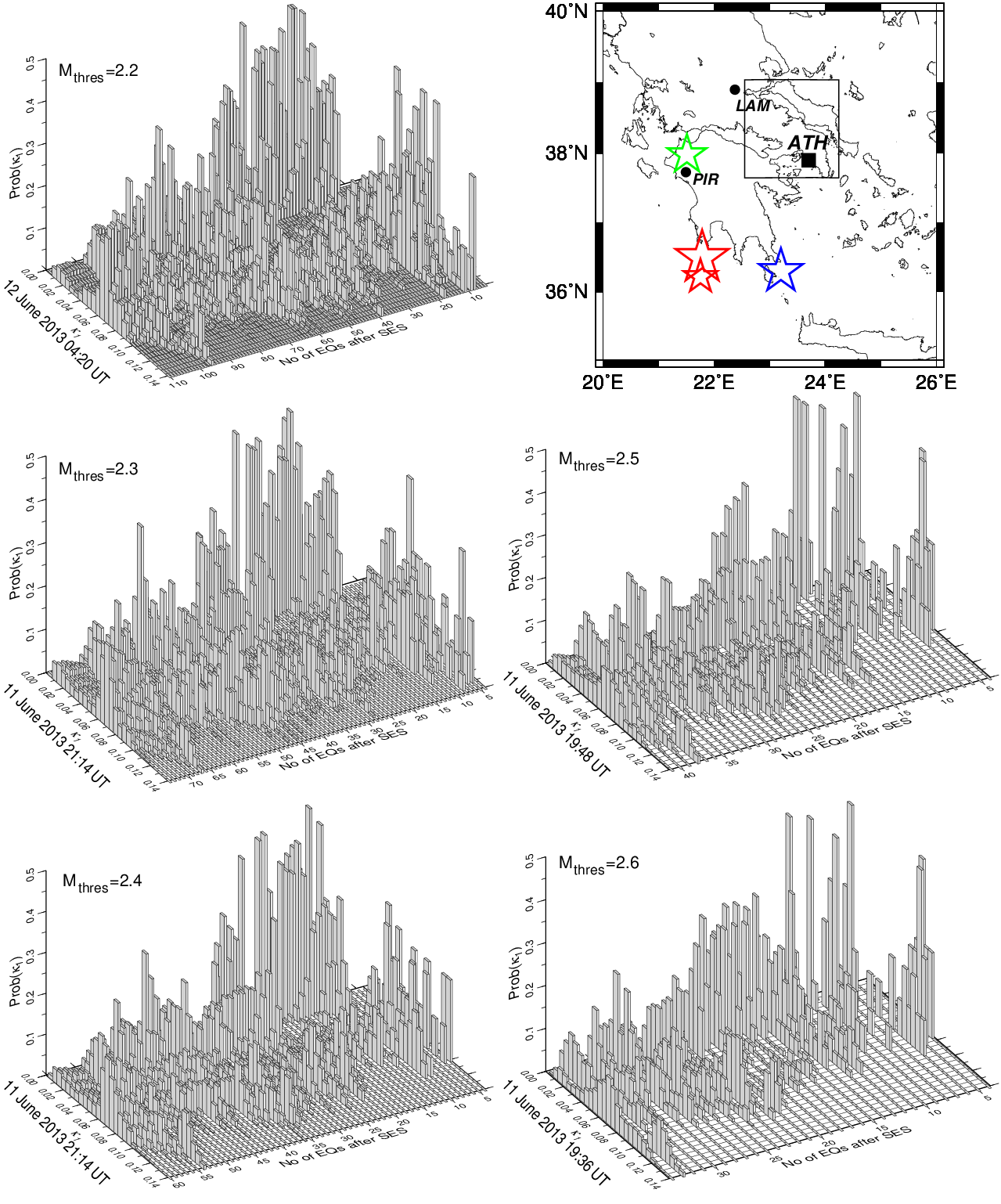}
\caption{(color online) The probability Prob$(\kappa_1)$ as it
results from the analysis of seismicity after the initiation of
the SES activity depicted in the upper panel of Fig.\ref{f1}(b)
within the rectangular area depicted in the map (uppermost right)
for various magnitude thresholds $M_{thres}$. The date and time of
the most recent earthquake considered into the calculation (upon
the occurrence of which Prob$(\kappa_1)$ maximized at $\kappa_1 =
0.070$) is written in each case.}\label{flam}
\end{figure*}

The fluctuations of $\kappa_1$, are studied by applying the
following procedure \cite{SPRINGER}. Taking an excerpt of a
seismic catalog comprising $W (\geq 100)$ successive events, we
start from the first EQ and calculate the first 35 $\kappa_1$
values for 6 to 40 consecutive EQs. Then we proceed to the second
EQ, and calculate again 35 values of $\kappa_1$ from the 7-th  to
the 41-st event. Thus, scanning event by event the whole excerpt
of $W$ earthquakes, we calculate the average value $\mu(\kappa_1)$
and the standard deviation $\sigma(\kappa_1)$ of the $\kappa_1$
values. The quantity
\begin{equation}\label{beta}
\beta \equiv \sigma(\kappa_1)/\mu(\kappa_1)
\end{equation}
is defined\cite{NEWEPL} as the variability $\beta$ of $\kappa_1$
for this excerpt of length $W$. In some occasions, as in the
present case, it is of prominent importance to know what happens
to the $\beta$ value until just before the occurrence of each EQ,
$e_i$, in the seismic catalog. We then calculate first the
$\kappa_1$ values using the {\em previous} $l$=6 to 40 consecutive
EQs. These 35 $\kappa_1$ values are associated with the EQ $e_i$,
but we clarify that EQ $e_i$ has not been employed for their
calculation. The $\beta$ value -corresponding to the EQ $e_i$- for
a natural time window length $W$ is computed using all the ($35
\times W$) $\kappa_1$ values associated with the EQs $e_{i-W+1}$
to $e_i$. The resulting value is denoted by $\beta_W$, where the
subscript $W$ shows the natural time window length, and the
corresponding minimum is designated by $\beta_{W,min}$.

\begin{widetext}
It is shown that the quantity $\beta$ when using $l$ consecutive
events is interrelated with the event's correlations through
 \begin{equation}\label{eqbeta}
 \beta=\frac{\sqrt{-\sum_{{\rm all} \phantom{x} {\rm pairs}} \left[ \left(\frac{m}{l}-\langle \chi \rangle_{\mathcal M} \right)^2 - \left(\frac{j}{l}-\langle \chi \rangle_{\mathcal M} \right)^2\right]^2
{\rm Cov} (p_j,p_m) -\left[ \sum_{{\rm all} \phantom{x} {\rm
pairs}} \frac{(j-m)^2}{l^2} {\rm Cov} (p_j,p_m)
\right]^2}}{\kappa_{1,{\mathcal M}}+\sum_{{\rm all} \phantom{x}
{\rm pairs}} \frac{(j-m)^2}{l^2} {\rm Cov} (p_j,p_m)},
 \end{equation}
where $\langle \chi \rangle_{\mathcal M}$  and
$\kappa_{1,{\mathcal M}}$ correspond to the average value of $\chi$
and $\kappa_1$, respectively, obtained when substituting for $p_k$ the average -within an excerpt of $W$ events- values  $\mu_k$
of $p_k$ ; the symbol ${\rm Cov} (p_j,p_m)$ stands
for covariance, i.e., the average value of $(p_j-\mu_j)(p_m-\mu_m)$ within the
excerpt of $W$ events. The details of the derivation of
Eq.(\ref{eqbeta}) are given in  the Appendix.
\end{widetext}

The selection of the $W$ value used for the purpose of our study
is of crucial importance. It is taken equal to the number of the
events that would occur in a few months, or so, in view of the
following: Low frequency ($\leq 1$ Hz) electric signals, termed
Seismic Electric Signals (SES), appear before earthquakes
\cite{VAR84A,VAR84B}. They are emitted from the future focal
region \cite{VARBOOK} (see also Ref. \cite{VAR93}) when in the
focal region the stress reaches a {\em critical} value
$\sigma_{cr}$, and then a {\em cooperative} orientation of the
electric dipoles occurs. This leads to the emission of a transient
electric signal that constitutes an SES. Several such signals
within a short time  are termed SES activity
\cite{VAR91,VAR93,NAT03A,NAT03B}. For example, the three lower
channels in Fig.\ref{f1}(b) show three SES activities that
preceded major earthquakes in western, southwestern and southern
Greece, respectively, as depicted in the map of Fig.\ref{f1}(a).
(Only for earthquakes of magnitude 6.0 or larger the SES
activities are publicized, see p.102 of Ref.\cite{NEWBOOK}.)
Furthermore, for the sake of comparison, the upper channel in
Fig.\ref{f1}(b) shows a recent SES activity initiated on 8 January
2013 at a station labelled LAM in Fig. \ref{f1}(a)  in central
Greece (cf. On 4 June 2013 an $M_L$4.3 earthquake occurred at
37.98$^o$N24.01$^o$E, i.e., around 20km E of Athens (ATH), which
is consistent with the earlier finding\cite{NEWBOOK} in 1999 that
LAM station is sensitive to seismic areas close to ATH. The
analysis, which was made by following Ref.\cite{SAR08}, continued
after this earthquake and showed the following results: The
probability Prob$(\kappa_1)$ of the $\kappa_1$ values of
seismicity in the area $N_{37.7}^{39.0}E_{22.6}^{24.2}$ maximized
at $\kappa_1=0.070$ on 11 \& 12 June 2013 exhibiting magnitude
threshold invariance for magnitudes in the range $M_{thres}=2.2$
to 2.6, see Fig.\ref{flam}, although the completeness of the
seismic catalog for such small magnitude thresholds is unclear.
These results seem to suggest that the system approaches the
critical point and conforms with the fact that a sequence of
additional SES activities were recorded at LAM from 31 March to 11
April 2013, see also Refs.\cite{neo,neo2,neo3,neo4,neo5,neo6,neo7}). The following
important fact has just been identified \cite{TECTO12}: At the
initiation of an SES activity, which usually occurs a few months
(with an upper limit of around 5 months) before a major EQ, a
clearly {detectable} change in seismicity appears, manifested by a
minimum $\beta_{W,min}$ in the fluctuations of the order parameter
of seismicity. Hence, in the case that geoelectrical data are
lacking, once we identify the date of $\beta_{W,min}$ (by
analyzing solely seismic data) this reveals also the date of an
SES activity that would have been recorded.

\begin{figure*}
\includegraphics[scale=0.5]{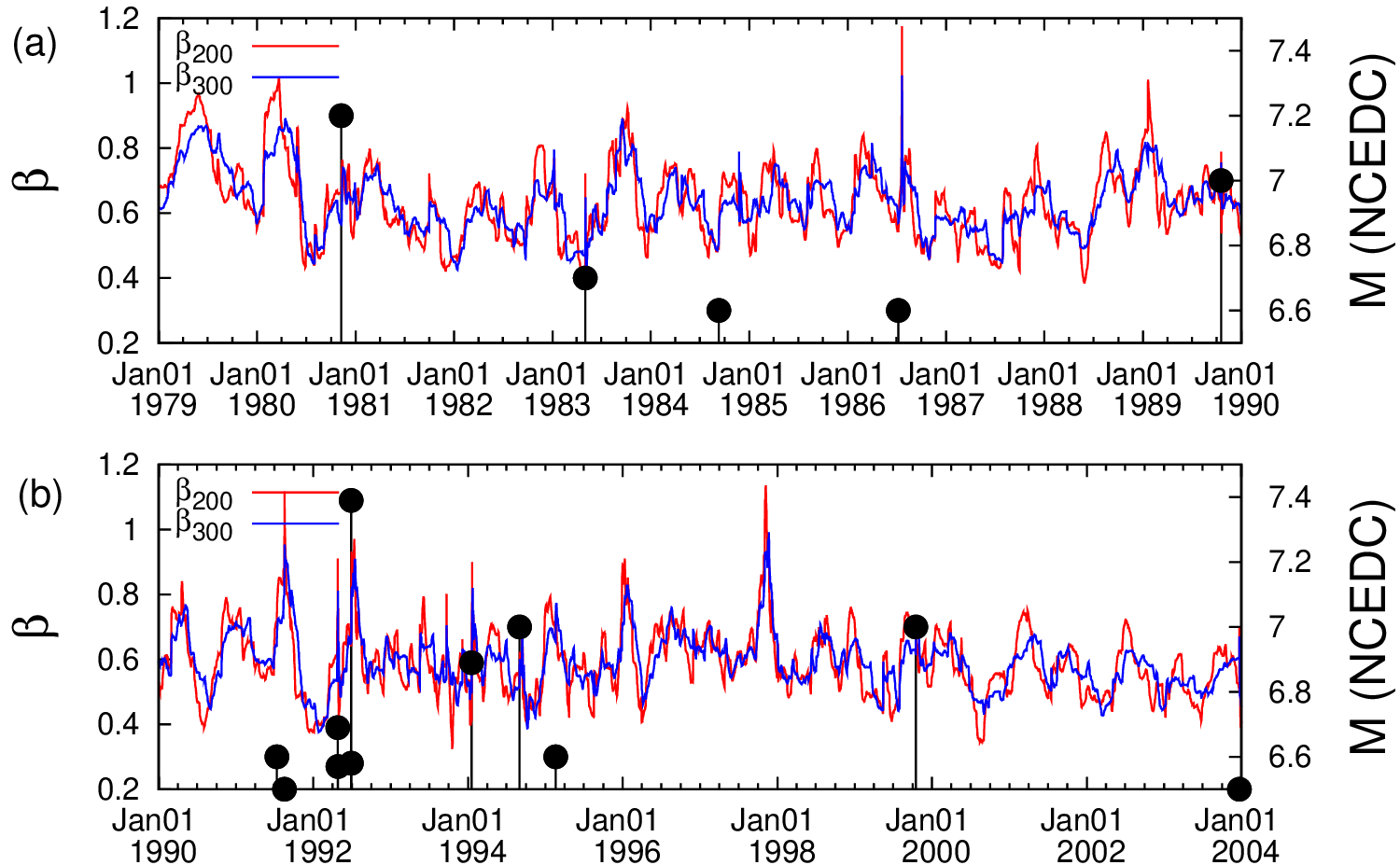}
\caption{(color online) The variability $\beta$ of $\kappa_1$
(left scale) plotted versus the conventional time for a natural
time window of length $W=$200 events (red) and $W=$300 events
(blue) during the period: (a) 1 January 1979 to 1 January 1990 and
(b) 1 January 1990 to 1 January 2004. The earthquakes with $M\geq
6.5$ (right scale) are shown with vertical bars ending at solid
circles.}\label{f2}
\end{figure*}

\begin{figure}[t]
\vspace*{2mm}
\begin{center}
\includegraphics[angle=-90,width=8.5cm]{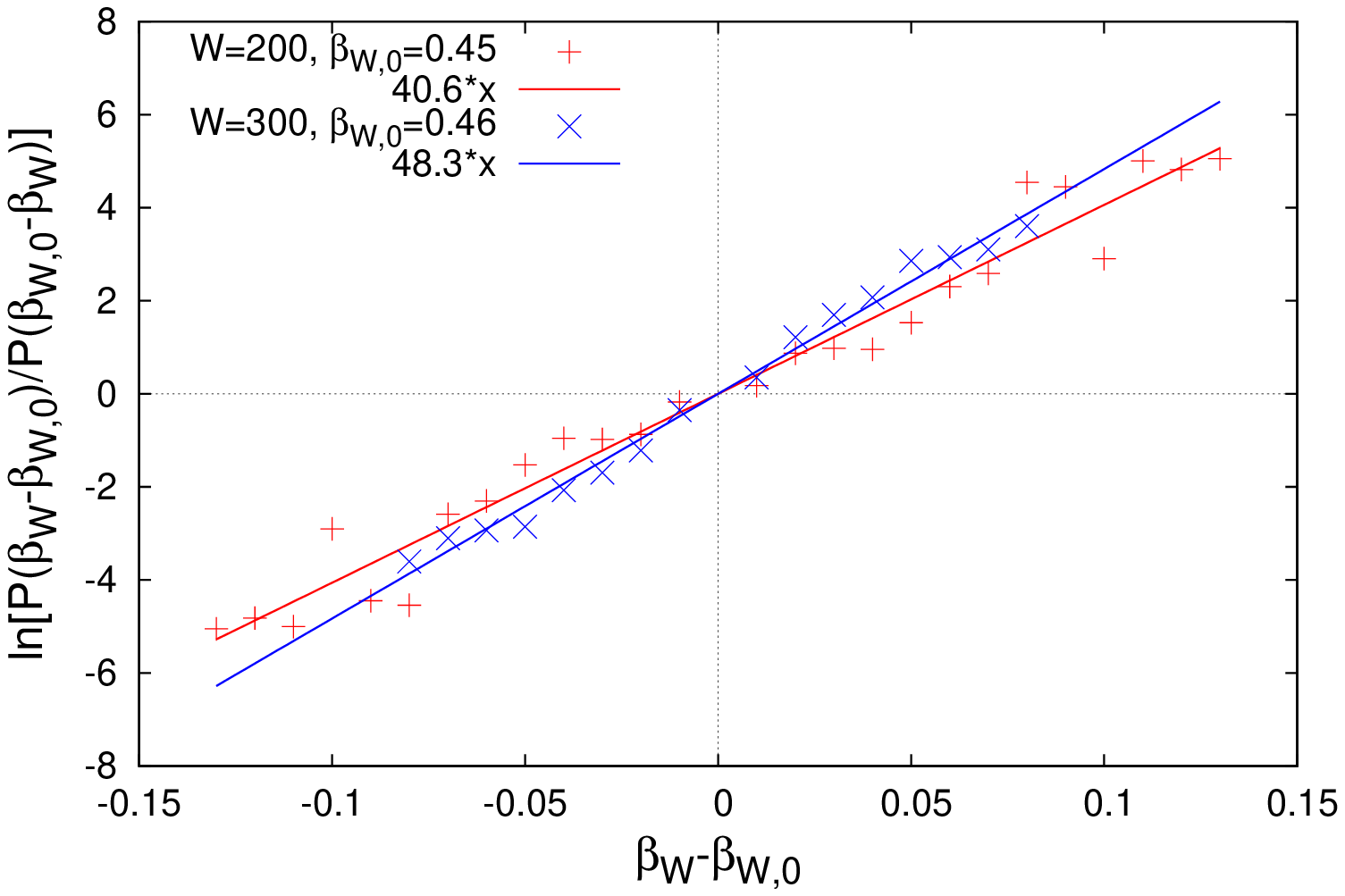}
\end{center}
\caption{\label{fig3}Application of Eq.(\ref{neweq}) for the
experimentally determined $\beta_W$ for $W=200$ (red) and 300
(blue). The threshold values $\beta_{200,0}=0.45$ and $\beta_{300,0}=0.46$
are deduced from the maximization of the linear correlation
coefficient (Pearson's) $r$, thus pointing\cite{numrec} to
optimal linearity. }
\end{figure}

\begin{table*}
\caption{The minimum DFA exponent $\alpha_{min}$ along with the
values of the minima observed for the variability $\beta$ together
with the dates of their observation in parentheses before all
major EQs in California with $M\geq 7.0$ within $N_{31.7}^{45.7}
W_{127.5}^{112.1}$ during the period 1979-2003. The $M6.9$
Northridge earthquake is also added in italics. The lead time
$\Delta t$ for each case, estimated from the difference in the
dates between the EQ occurrence and the appearance of
$\beta_{300,min}$ is shown in the last column. The values for
$\beta_{300,min}$ and $\alpha_{min}$ are taken from Ref.
\cite{EPL12}} \label{tab1}
\begin{tabular}{ccccccc}
\hline
EQ Date & EQ Name & $M$  & $\beta_{300,min}$ & $\beta_{200,min}$ & $\alpha_{{\rm min}}$ & $\Delta t$ (months) \\
 & epicenter &  &  (date) &  (date) & (date)  &  \\
\hline
1980-11-08 & Eureka & 7.2 & 0.444 & 0.432 & 0.445 & $\approx$ 3 \\
&N41.08$^o$W124.62$^o$ & & (1980-08-01) & (1980-06-28) & (1980-08-01) &\\
1989-10-18 & Loma Prieta & 7.0 & - & - & - & - \\
& N37.04$^o$W121.88$^o$ & & & &  &\\
1992-06-28 & Landers & 7.4 & 0.378 & 0.377 & 0.383 & $\leq$ 5 \\
&N34.19$^o$W116.46$^o$ & &(1992-01-28) & (1992-01-03) & (1992-02-02) &\\
{\em 1994-01-17} & {\em Northridge} & {\em 6.9} & {\em 0.459} & {\em 0.324} & {\em 0.431} & {\em $\approx$ 2} \\
& {\em N34.23$^o$W118.55$^o$ }& &{\em (1993-11-14)} & {\em (1993-10-18)} & {\em (1993-11-14)} &\\
1994-09-01 & Mendocino & 7.0 & 0.472 & 0.474 & 0.458 & $\approx$ 1 \\
& N40.41$^o$W126.30$^o$ & &(1994-08-01) & (1994-07-11) & (1994-08-09) &\\
1999-10-16 & Hector Mine & 7.0 & 0.444 & 0.432 & 0.422 & $\approx$ 5 \\
& N34.60$^o$W116.34$^o$ & &(1999-05-14) & (1999-05-14) & (1999-05-15) &\\
\\
{ Fluctuation theorem} & { and natural time analysis}  &  & { 0.46}  & { 0.45} &  &  \\
\hline
\end{tabular}
\end{table*}

Along these lines, Table \ref{tab1} shows the dates of the minima
$\beta_{W,min}$ of seismicity  before major mainshocks in California during the 25 year period 1 January 1979 to 1 January 2004. We used the
United States Geological Survey Northern California Seismic
Network catalog available from the Northern California Earthquake
Data Center, at the http address: {\tt
www.ncedc.org/ncedc/catalog-search.hmtl}, hereafter called NCEDC.
The seismic moment $M_0$, which is proportional to the energy
release during an earthquake and hence to the quantity $Q_k$ used
in natural time analysis, is calculated \cite{SPRINGER} from the
relation $\log_{10}(M_0)=1.5M+$const, where the earthquake
magnitudes reported in this catalog are labelled with $M$. The
earthquakes with $M \geq 2.5$ reported by NCEDC, within the area
${\rm N}_{31.7}^{45.7} {\rm W}_{127.5}^{112.1}$ have been
considered. We have on average $\sim 10^2$ EQs per month since
31832 earthquakes occurred for the 25 year period from 1 January
1979 to 1 January 2004. Thus, we adopted natural time window
lengths $W=$200 and $W=$300.

The results of this analysis are depicted in Fig. \ref{f2}(a),(b)
where we plot the variability $\beta$ (in red for $W=200$ and in
blue for $W=300$) versus the conventional time for the periods (a)
1 January 1979 to 1 January 1990 and (b) 1 January 1990 to 1
January 2004. An inspection of these results lead to the
$\beta_{W,min}$ values inserted in Table \ref{tab1}: In five out
of the six mainshocks we find values of $\beta_{300,min}$ and
$\beta_{200,min}$ that appear 1 to 5 months before mainshocks. In
these five cases $\beta_{200,min}$ varies between 0.324 to 0.474
and $\beta_{300,min}$ between 0.378 and 0.472. We note that the
key criterion to distinguish the true precursory $\beta_{W,min}$
from the non precursory ones is the following \cite{EPL12}: The
minimum should be followed (before the occurrence of the
mainshock) by a period during which the exponent $\alpha$ of the
Detrended Fluctuation Analysis (DFA) \cite{PEN94} -calculated for
a length $W$=300 events in the magnitude time series- reaches a
minimum $\alpha_{min}$ slightly smaller than 0.5 (thus, indicating
anticorrelated behavior, but close to random) and then
$\beta_{200} > \beta_{300}$. This inequality means that when the
system approaches closer to the critical point -which is the case
when considering $W$=200 events compared to $W$=300 events- the
fluctuations of the order parameter become more intense.

We now proceed to the investigation of Eq.(\ref{eq1}) in the case
of seismicity and  analyze the statistical distribution of the
experimentally determined $\beta_{W}$ for $W=200$ or 300, which is
clearly path depended. The quantity $\beta_{W}$ can be considered
as an entropic measure (see Appendix), but its
sign is by definition always positive. Thus, in order to apply
Eq.(\ref{eq1}), we need to define a threshold value $\beta_{W,0}$
above which the entropy production may be considered positive
whereas when below negative. For this reason, we employ the
relation
\begin{equation}\label{neweq}
\frac{Pr\left(\beta_{W}-\beta_{W,0}\right)}{Pr\left(\beta_{W,0}-\beta_{W}\right)}=\exp
\left[\tau' \left(\beta_{W}-\beta_{W,0}\right) \right],
\end{equation}
which results from Eq.(\ref{eq1}) when considering
$A=\beta_{W}-\beta_{W,0}$ and experimentally determine
$Pr\left(\beta_{W}-\beta_{W,0}\right)$ by using bins of width
$\Delta \beta_W=$0.01. Figure \ref{fig3} depicts the natural
logarithm of the left hand side of Eq.(\ref{neweq}) as a function
of $\left(\beta_{W}-\beta_{W,0}\right)$ for $W=200$ and $W=300$.
In each case, the threshold $\beta_{W,0}$  is the one that
maximizes the linear correlation coefficient (Pearson's)  $r$,
thus pointing\cite{numrec} to optimal linearity. We find the
threshold values of $\beta_{200,0}=0.45$ and $\beta_{300,0}=0.46$.
  Moreover, the relative `time-scale'
$\tau'$, which corresponds to the slope of Fig.\ref{fig3}, lies in
the range 40.6 to 48.3, which is comparable with a scale of the
order of $l=$40 sequential events used in the calculation of
$\beta_W$.

Let us now compare the aforementioned threshold values
$\beta_{200,0}=0.45$ and $\beta_{300,0}=0.46$  with the
$\beta_{W,min}$ values identified before each mainshock in Table
\ref{tab1}. We find that except one mainshock, i.e., the Mendocino
EQ in 1994, the other four mainshocks (including the strongest in
Table \ref{tab1}) led to $\beta_{200,min}$ and $\beta_{300,min}$
values that are {\em lower} than $\beta_{200,0}$ and
$\beta_{300,0}$, respectively.

Thus, in summary, it may be considered that once the natural time
analysis  leads to an identification of precursory minima (i.e.,
$\beta_{200,min}$ and $\beta_{300,min}$) that are {\em lower}
compared to the threshold $\beta_{W,0}$ values determined from the combined
use of natural time analysis  with the fluctuation theorem, a forthcoming
major EQ is likely to occur.

\begin{figure*}
\includegraphics[scale=0.8]{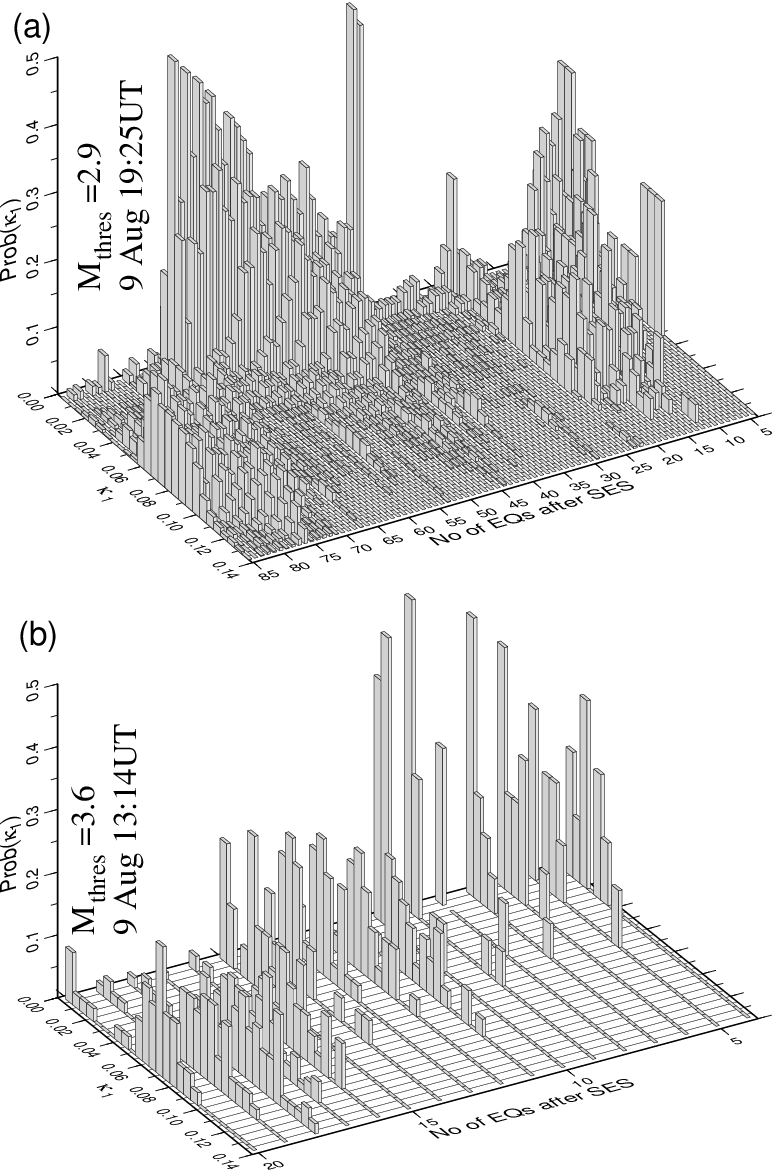}
\caption{The probability Prob$(\kappa_1)$ as it
results from the analysis of seismicity that occurred after the initiation (at 02:30 UT on 31 March 2013) of
the sequence of the additional SES activities from 31 March to 11 April 2013 (reported in the previous version of this manuscript on 13 June 2013)
within the rectangular area depicted in the map (uppermost right) of Fig.2
for various magnitude thresholds $M_{thres}$. The date and time of
the most recent earthquake considered into the calculation (upon
the occurrence of which Prob$(\kappa_1)$ maximized at $\kappa_1 =
0.070$) is written in each case.}\label{neofig}
\end{figure*}

\begin{figure}
\includegraphics{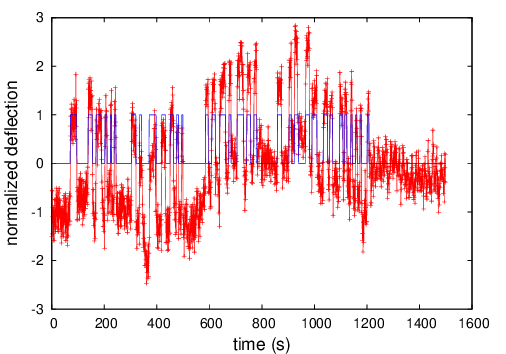}
\caption{The  SES activity recorded on 8 February 2014 at the station VOL (near Volos city, central Greece).
The normalized deflection versus time (lines with points) together with its dichotomous representation (lines without points).
The data start at 17:05 UTC. The analysis of this SES activity in natural time 
resulted in the following values $\kappa_1=0.076(3)$, $S=0.092(3)$,  and $S_{-}=0.076(3)$ which are compatible with those 
reported for SES activities (cf. $S$ and $S_-$ denote the entropy and the entropy under time reversal in natural time,
see Chapter 3 and Table 4.6 of Ref.\cite{SPRINGER}).}\label{VOLOS}
\end{figure}

\appendix*

\section{Interrelation of the variability $\beta$ with correlations when a (natural) time window of fixed length is sliding through a time series}
Here, we focus on the mean value $\mu \equiv \mathcal{E}
(\kappa_1)$ of $\kappa_1$ and  the corresponding standard
deviation $\sigma\equiv \sqrt{\mathcal{E}
\left[\kappa_1-\mathcal{E} (\kappa_1)  \right]^2}$ when a (natural
time) window of length $l$ is sliding through a time series of
$Q_k
> 0, k=1, 2,\ldots W$. Once these quantities have been evaluated,
the variability $\beta$ of $\kappa_1$ can then be estimated by
$\beta\equiv  \sigma /\mu$.

\subsection{The mean value $\mu \equiv \mathcal{E}
(\kappa_1)$ of $\kappa_1$}
 In a window of length $l$
starting at $k=k_0$, the quantities \begin{equation}
p_j(k_0)=\frac{Q_{k_0+j-1}}{\sum_{m=1}^l Q_{k_0+m-1}}, \phantom{x}
j=1,2,\dots,l\end{equation} representing the normalized energy are
obtained, which satisfy the necessary conditions
\begin{equation} \label{ap-1}
p_j(k_0)> 0,
\end{equation}
\begin{equation}
\label{ap0} \sum_{j=1}^l p_j(k_0)=1
\end{equation}
to be considered as point probabilities. We can then define as
usual\cite{NAT01,NAT02} the moments of the natural time
$\chi_j=j/l$ as $\langle \chi^q \rangle= \sum_{j=1}^l (j/l)^q
p_j(k_0)$ and hence
\begin{equation}\label{ap1}
  \kappa_1(k_0) = \sum_{j=1}^l \left( \frac{j}{l} \right)^2 p_j(k_0) - \left[ \sum_{j=1}^l \frac{j}{l}  p_j(k_0) \right]^2.
\end{equation}
Note that $\kappa_1$ is a non-linear functional of $\{p_j \}$.

\begin{widetext}
Let us consider the average value $\mu_j$ of $p_j$ obtained when
the (natural time) window of length $l$ slides through a time
series of $Q_k > 0, k=1, 2,\ldots W$, i.e., we
have\cite{SARCHRIS12A}
\begin{equation}\label{muj}
\mu_j\equiv\mathcal{E}(p_j)=\frac{1}{W-l+1}\sum_{k_0=1}^{W-l+1}p_j(k_0)=\frac{1}{W-l+1}\sum_{k_0=1}^{W-l+1}
\frac{Q_{k_0+j-1}}{\sum_{m=1}^l Q_{k_0+m-1}}.
\end{equation} It is obvious that the definition of Eq.(\ref{muj}) is consistent with Eq.(\ref{ap0}), thus  we have \begin{equation} \label{muj2}
\sum_{j=1}^l \mu_j=1.
\end{equation}
Similarly for the second order moments of $p_j$, one can
estimate\cite{SARCHRIS12A} the variance of $p_j$ by
\begin{equation}
{\rm Var}(p_j)\equiv \mathcal{E} \left[  \left( p_j-\mu_j\right)^2
\right] =\frac{1}{W-l+1}\sum_{k_0=1}^{W-l+1}
\left(\frac{Q_{k_0+j-1}}{\sum_{m=1}^l Q_{k_0+m-1}}-\mu_j \right)^2
\end{equation}
as well as the covariance
\begin{eqnarray}
{\rm Cov}(p_j,p_i)&\equiv& \mathcal{E} \left[  \left( p_j-\mu_j\right) \left( p_i-\mu_i\right) \right] \nonumber \\
&=&\frac{1}{W-l+1}\sum_{k_0=1}^{W-l+1}
\left(\frac{Q_{k_0+j-1}}{\sum_{m=1}^l Q_{k_0+m-1}}-\mu_j \right)
\left(\frac{Q_{k_0+i-1}}{\sum_{m=1}^l Q_{k_0+m-1}}-\mu_i \right).
\end{eqnarray}
In view of Eqs.(\ref{ap-1}) and(\ref{ap0}), the quantities
$\mu_j$, ${\rm Var}(p_j)$ and ${\rm Cov} (p_j,p_m)$ are always
finite irrespective of the existence  of heavy tails in $Q_k$
which is for example the case of seismicity. Moreover, for the
purpose of our calculations the relation between the variance of
$p_j$, ${\rm Var}(p_j)$, and the covariance of $p_j$ and $p_m$,
${\rm Cov} (p_j,p_m)$,  is important.  Equations (\ref{ap0}) and
(\ref{muj2}) lead to
\begin{equation}
p_j-\mu_j=\sum_{m\neq j}(\mu_m-p_m),
\end{equation}
which when multiplied by $(p_j-\mu_j)$ and averaged (cf.
$\hat{\mathcal{E}}\equiv \frac{1}{W-l+1}\sum_{k_0=1}^{W-l+1})$
results in
\begin{equation}
{\rm Var}(p_j)=-\sum_{m\neq j} {\rm Cov} (p_j,p_m).
\label{a2natimebcg}
\end{equation}

We now turn to the evaluation of the mean value $\mu$ of
$\kappa_1$ obtained when the (natural time) window of length $l$
slides through a time series of $Q_k > 0, k=1, 2,\ldots W$,
\begin{equation}\mu\equiv\mathcal{E} (\kappa_1)=\frac{1}{W-l+1}\sum_{k_0=1}^{W-l+1}\kappa_1(k_0),\end{equation}
by studying its difference from the one that corresponds to the
time series of the averages ${\mathcal M}=\{ \mu_k \}$ which is
labelled $\kappa_{1,{\mathcal M}}$,
\begin{equation}\label{ap1nikos}
  \kappa_{1,{\mathcal M}} = \sum_{j=1}^l \left( \frac{j}{l} \right)^2 \mu_j - \left[ \sum_{j=1}^l \frac{j}{l}  \mu_j \right]^2.
\end{equation}
 Hence,
\begin{equation}
\label{a4natimebcg} \mu-\kappa_{1,{\mathcal
M}}=\frac{1}{W-l+1}\sum_{k_0=1}^{W-l+1} \left\{ \sum_{m=1}^l
\frac{m^2}{l^2} \left[ p_m(k_0)-\mu_m  \right] - \left[
\sum_{m=1}^l \frac{m}{l}p_m(k_0)\right]^2+ \left( \sum_{m=1}^l
\frac{m}{l}\mu_m \right)^2               \right\}.
\end{equation}
In view of the definition of $\mu_m$, the first term in square
brackets in the right hand side of Eq.(\ref{a4natimebcg})
vanishes, whereas the latter two terms reduce to the opposite of
the variance of
\begin{equation}
\langle \chi \rangle_{\mathcal M}=\sum_{m=1}^l \frac{m}{l} \mu_m,
\end{equation}
leading to
\begin{equation}
\label{a5natimebcg} \mu-\kappa_{1,{\mathcal
M}}=-\frac{1}{W-l+1}\sum_{k_0=1}^{W-l+1}
 \left\{ \sum_{m=1}^l \frac{m}{l}\left[ p_m(k_0)-\mu_m \right]
\right\}^2  .
\end{equation}
{Expanding the term within the curly brackets and interchanging
the summations, we get}
\begin{equation}
\label{a6} \kappa_{1,{\mathcal M}}-\mu=\sum_{m=1}^l
\frac{m^2}{l^2} {\rm Var}(p_m)+2 \sum_{j=1}^{l-1} \sum_{m=j+1}^{l}
\frac{jm}{l^2} {\rm Cov} (p_j,p_m).
\end{equation}
which, upon using Eq.(\ref{a2natimebcg}), leads to
\begin{equation}
\label{a7} \mu-\kappa_{1,{\mathcal M}}=\sum_{j=1}^{l-1}
\sum_{m=j+1}^{l} \frac{(j-m)^2}{l^2} {\rm Cov}
(p_j,p_m)=\frac{1}{2} \sum_{j=1}^{l}\sum_{m=1}^{l}
\frac{(j-m)^2}{l^2} {\rm Cov} (p_j,p_m).
\end{equation}
The latter  relation turns to
\begin{equation}
\label{a8}\mu= \kappa_{1,{\mathcal M}}+\sum_{{\rm all} \phantom{x}
{\rm pairs}} \frac{(j-m)^2}{l^2} {\rm Cov} (p_j,p_m)
\end{equation}
where $\sum_{{\rm all pairs}}\equiv
\sum_{j=1}^{l-1}\sum_{m=j+1}^{l}$.

Equation (\ref{a8})  shows that the mean value $\mu$ itself is a
measure of the correlations between successive earthquake
magnitudes. The practical use of this equation, however, in order
to estimate the strength of these correlations between seismic
events requires\cite{NAT06B,NAT09B,SAR11,SPRINGER,SARCHRIS12A} the
construction of a large number of shuffled copies of the original
earthquake catalog and a comparison of $\mu$ with the relevant
distribution obtained from the shuffled copies. Obviously, this
task becomes cumbersome when the (natural time) window of length
$l$ is sliding through a long time series of $Q_k$.

When $Q_k$ {\em are independent and identically distributed
positive random variables}, Eq.(\ref{a8})
results\cite{NAT06B,SPRINGER} in
\begin{equation}
\label{a9i} \mu= \kappa_u \left( 1-\frac{1}{l^2} \right)-\kappa_u
(l+1){\rm Var}(p),
\end{equation}
where $\kappa_u=1/12$ -corresponding to the $\kappa_1$ value for
the uniform distribution- and ${\rm Var}(p)$ the variance of any
$p_j$.

\subsection{The standard deviation $\sigma$ of the $\kappa_1$ values}
Let us now investigate the standard deviation $\sigma$ of the
$\kappa_1$ values obtained when the (natural time) window of
length $l$ slides through a time series of $Q_k$. This is obtained
from the variance
\begin{equation}\label{a9}
\sigma^2={\rm Var}\left({\kappa_1}\right)\equiv\mathcal{E} \left[
\left( \kappa_1-\mu
\right)^2\right]=\frac{1}{W-l+1}\sum_{k_0=1}^{W-l+1}\left[
\kappa_1(k_0)-\mu \right]^2.\end{equation} Numerically, the above
quantity can be evaluated almost as easily  as $\mu$ when
$\kappa_1(k_0)$ are available.

In order to obtain an analytical expression, by inserting
Eq.(\ref{a8}) into (\ref{a9}), we obtain
\begin{eqnarray}
\sigma^2&=&\frac{1}{W-l+1}\sum_{k_0=1}^{W-l+1}\left\{ \sum_{m=1}^l
\frac{m^2}{l^2} \left[ p_m(k_0)-\mu_m  \right] -
\left[ \sum_{m=1}^l \frac{m}{l}p_m(k_0)\right]^2 \right. \nonumber \\
&+& \left. \left( \sum_{m=1}^l \frac{m}{l}\mu_m \right)^2
-\sum_{{\rm all} \phantom{x} {\rm pairs}} \frac{(j-m)^2}{l^2} {\rm
Cov} (p_j,p_m)            \right\}^2.
\end{eqnarray}
Rearranging the terms
\begin{eqnarray}
\left[ \sum_{m=1}^l \frac{m}{l}p_m(k_0)\right]^2&-&\left(
\sum_{m=1}^l
\frac{m}{l}\mu_m \right)^2 =\left\{ \sum_{m=1}^l \frac{m}{l}\left[ p_m(k_0)-\mu_m\right]\right\} \left\{ \sum_{m=1}^l \frac{m}{l}\left[ p_m(k_0)+\mu_m\right]\right\} \nonumber \\
&=& \left\{ \sum_{m=1}^l \frac{m}{l}\left[
p_m(k_0)-\mu_m\right]\right\}^2+2 \langle \chi \rangle_{\mathcal
M} \left\{ \sum_{m=1}^l \frac{m}{l}\left[
p_m(k_0)-\mu_m\right]\right\}
\end{eqnarray}
we get
\begin{eqnarray}\label{mus}
\sigma^2&=&\frac{1}{W-l+1}\sum_{k_0=1}^{W-l+1}\left[ \sum_{m=1}^l
\left( \frac{m^2}{l^2} -2  \langle \chi \rangle_{\mathcal M}
\frac{m}{l} \right) \left[ p_m(k_0)-\mu_m  \right] -
\left\{ \sum_{m=1}^l \frac{m}{l}\left[ p_m(k_0)-\mu_m \right] \right\}^2 \right. \nonumber \\
&-& \left. \sum_{{\rm all} \phantom{x} {\rm pairs}}
\frac{(j-m)^2}{l^2} {\rm Cov} (p_j,p_m)            \right]^2.
\end{eqnarray}
Upon expanding the square over the square brackets in
Eq.(\ref{mus}) we obtain six terms:
\begin{subequations}
\begin{align}
\sigma^2&=\frac{1}{W-l+1}\sum_{k_0=1}^{W-l+1}\left\{ \sum_{m=1}^l \left( \frac{m^2}{l^2} -2  \langle \chi \rangle_{\mathcal M} \frac{m}{l} \right) \left[ p_m(k_0)-\mu_m  \right] \right\}^2 \label{term1}\\
&- \frac{2}{W-l+1}\sum_{k_0=1}^{W-l+1}\left\{ \sum_{m=1}^l \left( \frac{m^2}{l^2} -2  \langle \chi \rangle_{\mathcal M} \frac{m}{l} \right) \left[ p_m(k_0)-\mu_m  \right] \right\} \left\{ \sum_{m=1}^l \frac{m}{l}\left[ p_m(k_0)-\mu_m \right] \right\}^2 \label{term2} \\
&- \left[ \sum_{{\rm all} \phantom{x} {\rm pairs}}
\frac{(j-m)^2}{l^2}
{\rm Cov} (p_j,p_m)  \right]\frac{2}{W-l+1}\sum_{k_0=1}^{W-l+1}\left\{ \sum_{m=1}^l \left( \frac{m^2}{l^2} -2  \langle \chi \rangle_{\mathcal M} \frac{m}{l} \right) \left[ p_m(k_0)-\mu_m  \right] \right\}  \label{term3}\\
&+\frac{1}{W-l+1}\sum_{k_0=1}^{W-l+1}\left\{ \sum_{m=1}^l \frac{m}{l}\left[ p_m(k_0)-\mu_m \right] \right\}^4 \label{term4}\\
&+\left[ \sum_{{\rm all} \phantom{x} {\rm pairs}}
\frac{(j-m)^2}{l^2}
{\rm Cov} (p_j,p_m)  \right] \frac{2}{W-l+1}\sum_{k_0=1}^{W-l+1}\left\{ \sum_{m=1}^l \frac{m}{l}\left[ p_m(k_0)-\mu_m \right] \right\}^2 \label{term5}\\
&+\left[ \sum_{{\rm all} \phantom{x} {\rm pairs}}
\frac{(j-m)^2}{l^2} {\rm Cov} (p_j,p_m)  \right]^2. \label{term6}
\end{align}
\end{subequations}
The following comments are in  order: First, the term in
(\ref{term3}) vanishes due to Eq.(\ref{muj}). Second, the terms in
(\ref{term2}) and (\ref{term4}) clearly depend on moment
correlations higher than the second, thus they should be neglected
when restricting ourselves to second order correlations. Third,
the second term in (\ref{term5}) can be evaluated using
Eqs.(\ref{a5natimebcg}) and (\ref{a7}) leading to a partial
cancellation with the term in (\ref{term6}). Hence, {\em
restricting ourselves to second order correlations}, we finally
obtain
\begin{equation}\label{seconds}
\sigma^2=\frac{1}{W-l+1}\sum_{k_0=1}^{W-l+1}\left\{ \sum_{m=1}^l
\left( \frac{m^2}{l^2} -2  \langle \chi \rangle_{\mathcal M}
\frac{m}{l} \right) \left[ p_m(k_0)-\mu_m  \right]
\right\}^2-\left[ \sum_{{\rm all} \phantom{x} {\rm pairs}}
\frac{(j-m)^2}{l^2} {\rm Cov} (p_j,p_m)  \right]^2.
\end{equation}
The first term in Eq.(\ref{seconds}) can be evaluated by expanding
the square over the curly brackets and using
Eq.(\ref{a2natimebcg}) -in a way similar to Eqs.(\ref{a6}) and
(\ref{a7})- so that we obtain
\begin{equation}\label{seconds2}
\sigma^2= -\sum_{{\rm all} \phantom{x} {\rm pairs}} \left[
\left(\frac{m}{l}-\langle \chi \rangle_{\mathcal M} \right)^2 -
\left(\frac{j}{l}-\langle \chi \rangle_{\mathcal M}
\right)^2\right]^2 {\rm Cov} (p_j,p_m) -\left[ \sum_{{\rm all}
\phantom{x} {\rm pairs}} \frac{(j-m)^2}{l^2} {\rm Cov} (p_j,p_m)
\right]^2.
\end{equation}
Equation (\ref{seconds2}) reveals that $\sigma^2$ -like the mean
value $\mu$ in Eq.(\ref{a8})- is a measure of the  correlations,
but $\sigma^2$ is almost proportional (see also below) to these
correlations whereas in $\mu$ they appear as an additive term in
Eq.(\ref{a8}).

\subsection{The variability $\sigma/\mu$}
By combining Eqs.(\ref{a8}) and (\ref{seconds2}) we find:
 \begin{equation}\label{betaf}
 \beta=\frac{\sqrt{-\sum_{{\rm all} \phantom{x} {\rm pairs}} \left[ \left(\frac{m}{l}-\langle \chi \rangle_{\mathcal M} \right)^2 - \left(\frac{j}{l}-\langle \chi \rangle_{\mathcal M} \right)^2\right]^2
{\rm Cov} (p_j,p_m) -\left[ \sum_{{\rm all} \phantom{x} {\rm
pairs}} \frac{(j-m)^2}{l^2} {\rm Cov} (p_j,p_m)
\right]^2}}{\kappa_{1,{\mathcal M}}+\sum_{{\rm all} \phantom{x}
{\rm pairs}} \frac{(j-m)^2}{l^2} {\rm Cov} (p_j,p_m)}.
 \end{equation}
This equation, which is just Eq.(4) of the main text, provides in
general the interrelation between the variability $\beta$ and the
event correlations.

Additional insight on the physical meaning of $\sigma/\mu$ may be
obtained when adopting  the paradigm of the uniform
distribution\cite{NAT03A,NAT03B,SPRINGER} which corresponds to a
simple system operating at stationarity, i.e., when $Q_k$ are
independent and identically distributed positive random variables.
In this case, we have\cite{SPRINGER}
\begin{equation}
\mu_j=\frac{1}{l},
\end{equation}
\begin{equation}
\langle \chi \rangle_{\mathcal M}=\sum_{m=1}^l
\frac{m}{l^2}=\frac{1}{2}+\frac{1}{2l},
\end{equation}
and due to Eq.(\ref{a2natimebcg})
\begin{equation}
{\rm Cov}(p_j,p_m)=-\frac{{\rm Var}(p)}{(l-1)},\end{equation} thus
we obtain
\begin{equation}\label{seconds3}
\sigma^2= \frac{{\rm Var}(p)}{(l-1)}\left\{ \sum_{{\rm all}
\phantom{x} {\rm pairs}} \left[
\left(\frac{m}{l}-\frac{1}{2}-\frac{1}{2l} \right)^2 -
\left(\frac{j}{l}-\frac{1}{2}-\frac{1}{2l}\right)^2\right]^2
-\frac{{\rm Var}(p)}{(l-1)}\left[ \sum_{{\rm all} \phantom{x} {\rm
pairs}} \frac{(j-m)^2}{l^2}
 \right]^2 \right\}.
\end{equation}
For large $l$ the summations over all pairs can be effectively,
e.g. $l>10$, approximated by integrations
\begin{equation}
\sum_{{\rm all} \phantom{x} {\rm pairs}} \left[
\left(\frac{m}{l}-\frac{1}{2}-\frac{1}{2l} \right)^2 -
\left(\frac{j}{l}-\frac{1}{2}-\frac{1}{2l}\right)^2\right]^2\approx
\frac{l^2}{2}\int_0^1\int_0^1 \left[
\left(\chi-\frac{1}{2}\right)^2-
\left(\psi-\frac{1}{2}\right)^2\right]^2 d\chi
d\psi=\frac{l^2}{180},
\end{equation}
\begin{equation}\sum_{{\rm all} \phantom{x} {\rm pairs}} \frac{(j-m)^2}{l^2}
\approx \frac{l^2}{2}\int_0^1\int_0^1 \left(\chi-\psi\right)^2
d\chi d\psi=\frac{l^2}{12}
\end{equation}
\end{widetext}
Equation (\ref{seconds3}) simplifies to
\begin{equation}
\sigma^2\approx l{\rm Var}(p) \kappa_u^2  \left[\frac{4}{5}-l{\rm
Var}(p)\right],
\end{equation}
and Eq.(\ref{a9i}) becomes
\begin{equation}
\mu\approx \kappa_u \left[ 1-l{\rm Var}(p)\right].
\end{equation}
Thus, the variability simply results in
\begin{equation}\label{master}
\beta=\frac{\sigma}{\mu}=\sqrt{l{\rm Var}(p)} \left[
\frac{\sqrt{\frac{4}{5}-l{\rm Var}(p)}}{1-l{\rm Var}(p)} \right].
\end{equation}
{\em When $Q_k$ exhibit heavy tails} as in the case for
seismicity, the  quantity $l{\rm Var}(p)$ measures the intensity
of such tails and so does $\beta$. In other words, for  randomly
shuffled earthquake data or  earthquakes occurring with temporally
uncorrelated magnitudes, the variability $\beta$ is a measure of
the $b$-value of the Gutenberg-Richter law. Since real seismic
data may also exhibit temporal correlations between earthquake
magnitudes\cite{NAT06B,LIP07,NAT09B,SAR11,SPRINGER,SARCHRIS12A,LIP12},
the general expression of the variability obtained above (i.e.,
Eq.(\ref{betaf})) from Eqs.(\ref{a8}) and (\ref{seconds2}) {\em
captures both the effects of correlations and heavy-tails}.

{\em When $Q_k$ do not exhibit heavy tails,} which is not of
course the case of seismicity, the quantity $l{\rm Var}(p)$ is
simply related\cite{NAT04,SPRINGER} to the mean $\mu_0$ and the
standard deviation $\sigma_0$ of $Q_k$:
\begin{equation}
l{\rm Var} (p) = \frac{1}{l}\frac{\sigma_0^2}{\mu_0^2}.
\label{a3f}
\end{equation}
Assuming that $\sigma_0/\mu_0$ is of the order of unity, $l{\rm
Var} (p)$ becomes small compared to unity when $l>10$, and
Eq.(\ref{master}) becomes
\begin{equation}
\beta=\frac{\sigma}{\mu}=\frac{2}{\sqrt{5}}\frac{\sigma_0}{\mu_0}\left(
\frac{1}{\sqrt{l}}\right),
\end{equation}
i.e., the variability of $\kappa_1$ is directly  proportional to
the variability of the data $Q_k$. Note that the same holds for
the standard deviation of the natural time
entropy\cite{NAT04,SPRINGER} $S$ as well as for change $\Delta S$
of the entropy  in natural time under time
reversal\cite{NAT07,SPRINGER} (cf. for the analysis in natural
time under time reversal, see also Refs.\cite{NAT06A} and
\cite{NAT08}). Thus, in this case, one could alternatively
 view $\beta$ as an entropic measure.



\end{document}